\RequirePackage{fix-cm}
\documentclass[smallextended]{svjour3}       
\smartqed  
\usepackage{graphicx}
\begin{document}

\title{Ethanol chemisorption on core-shell Pt-nanoparticles: an {\em ab-initio} study
}


\author{Vagner A. Rigo         \and
        Caetano R. Miranda     \and
        Francesca Baletto       
}


\institute{Vagner A. Rigo \at
              Universidade Tecnol\'ogica Federal do Paran\'a (UTFPR), Corn\'elio Proc\'opio, Brazil \\
              \email{vagnerrigo@utfpr.edu.br}           
           \and
           Caetano R. Miranda \at
              Instituto de F\'isica, Universidade de Sao Paulo (IF-USP), Sao Paulo, Brazil \\
              \email{cmiranda@if.usp.br}           
           \and
           Francesca Baletto \at
              King's College, Physics Department, Strand WC2R 2LS, London, United Kingdom \\
              \email{francesca.baletto@kcl.ac.uk}           
}

\date{Received: date / Accepted: date}

\maketitle

\begin{abstract}
By means of {\em ab-initio} calculations, we have investigated the \linebreak \mbox{chemisorption} paroperties of ethanol onto segregating binary nanoalloys. 
We select nanostructures with icosahedral shape of 55 atoms with a Pt outermost layer over a M core with M=Ag,Pd,Ni. With respect to nanofilms with equivalent 
composition, there is an increse of the ethanol binding energy. This is not merely due to observed shortening of the Pt-O distance but 
depends on the nanoparticle distortion after ethanol adsorption. This geometrical distortion within the nanoparticle can be interpreted 
as a radial breathing, which is sensitive to the adsortion site, identified by the O-anchor point and the relative positions of the ethyl group.
More interestingly, being core-dependent -larger in Pd@Pt and smaller in Ni@Pt-, it relates to an effective electron transfer from ethanol 
and the M-core towards the Pt-shell. On the view of this new analysis, Pd@Pt nanoalloys show the most promissing features for ethanol oxidation.
%
\end{abstract}

\section{Introduction}
\label{intro}
Fuel cells can convert fuel into electricity with high efficiency, low noise, and emission rates \cite{An2015,Badwal2015,Sun2014}. 
Among these devices, direct ethanol fuel cells (DEFC) are particularly appealing since ethanol is a renewable fuel, can be produced
from a variety of different ways, is nontoxic, and for being liquid at ambient conditions, can take advantage of the existing fuel 
distribution network \cite{An2015,Badwal2015}. 

Indeed, in DEFCs, ethanol chemical energy is converted to electricity using a simple operation setup \cite{An2015,Badwal2015}, without the necessity to produce hydrogen 
first \cite{An2015,Badwal2015,Sun2014,Wu2016,Verna2007,An2014,An2013}. 
However, currently, the conversion reaction in the DEFCs stops at acetaldehyde and acetic acid, before to fully reach the ethanol 
oxidation to CO$_2$, even employing the state-of-the-art catalysts \cite{An2015,Sun2014}. Considering a 1:1 mole fraction of 
acetaldehyde and acetic acid products, 3 electrons are delivered per ethanol molecule, instead of the nominal 12 e$^{-}$ \cite{An2015}. 
On alkaline fuel cells, the ethanol oxidation reaction (EOR) kinetics at the anode is improved by adding a basic solution, 
e.g. KOH or NaOH, to ethanol \cite{An2015}. 
Even so, the reaction is limited by the formation of acetic acid, delivering only 4 e$^{-}$ per ethanol. 
Recently many efforts have been done on the search for promising catalytic materials to enhance ethanol 
kinetics \cite{An2015,Badwal2015,Sun2014,Wu2016,Verna2007,An2014,An2013,An2012,Antoniassi2016,Tereshchuk2012,Kowal2016,Tereshchuk2013}.

Better performance is observed in Pt-based binary or ternary alloys against simple compounds \cite{Engelbrekt2016,Chen2017}. 
{\it Ab-initio} studies on Au/Pt, Pd/Pt, Au/Pd, Pd/Au, Pt/Au (111) nanofilms show that the binding energy of ethanol on dealloyed 
Pt-monolayer above various metallic substrates is similar to the energy on the bare Pt(111) \cite{Pereira2014}. Other strategies 
try to downsize Pt-based catalysts to nanoscale, where recent experimental studies showed that Pt clusters could improve the 
selectivity to C-C cleaving and CO$_2$ production along EOR \cite{Antoniassi2016,Sun2015,Guan2009}.

The understanding of ethanol chemisorption on metallic clusters is still relatively poor and the studies are limited to a few cases, 
with core-shell systems often taken as promising candidates for enhancing EOR \cite{Du2014,Xie2016}. The magic size at which mass 
and specific activity peak can be optimized is still under debate. However, it is expected that Pt-nanoparticles of 2.5-2.6 nm 
enhance the specific activity as they balance the structural stability versus oxophilicity effects of the Pt surface \cite{Perez2011}.

Nevertheless, the atomistic details are still not fully understood yet, especially addressing the role of different chemical compositions 
and the variety of active sites of these nanosystems \cite{Davis2015}. From a modeling point of view, using density functional theory calculations with van der Waals corrected, 
Zibordi-Besse and co-workers reported that the ethanol adsorption on an icosahedron 
of 13-atoms moves from Ag, to Au, to Pd, to Cu, to Pt, then Ni, but the icosahedral symmetry is lost in both Au and Pt nanoclusters \cite{ZB2016}.

In this work, {\it ab-initio} simulations have been used to evaluate ethanol adsorption on 55-atoms binary nanoalloys (NAs), with a Pt 
outer shell over a Ag, Pd, or Ni core. We select an icosahedral shape, to limit the study to closed-shell geometry and for being 
representative among Pt-structural motifs in subnano regime \cite{Sandoval2017,Baletto2005,Baletto2002,Johnston2008}.
The proposed electronic and structural analysis show that all the considered NAs present a non-homogeneous radial breathing upon ethanol 
adsorption, but the intensity varies according to the metal-core, stronger in Pd@Pt weaker in Ni@Pt. At the same time, a distortion 
within ethanol is observed with C-C and O-C more distorted in Pd@Pt and less in Ni@Pt. A linear relationship between the Pt-oxygen distance 
and the nanoparticle distortion is proposed to describe the system energetics. As a general result, Pd@Pt icosahedra shows adsorption properties 
favourable to EOR, compared the nanofilm with equivalent composition \cite{Pereira2014}. This result can be further exploited in the design 
of more efficient nanocatalysts, with the aim to overcome the high cost of platinum on fuel cells applications. 

\section{Models and Methods}
\label{MM}

The spin-polarized total-energy density functional theory (DFT) calculations \cite{Hohenberg1964,Kohn1965} are performed using the Vienna Ab-initio
Simulation Package (VASP) \cite{Kresse1993,Kresse1996}. We employ the generalized gradient approximation of Perdew, Burke, and Ernzerhof
(PBE) \cite{PBE1996} for the exchange-correlation functional, the projector-augmented wave method \cite{Blochl1994,Kresse1999} for
atomic potentials, and Gamma-point calculations for Brillouin-zone sampling.
A plane-wave energy cutoff of 450 eV is used in all systems, and a vacuum slab of at least 12 \AA{} is adopted, following convergence analysis. The van
der Waals dispersion interaction is described using the Grimme formulation \cite{Grimme2010}, standardized named as PBE+D3. Geometries are
optimized using conjugated-gradient method until forces on atoms were lower than 0.03 eV/\AA{}. Atomic charges are obtained through Bader
analysis \cite{Tang2009}.

We consider three core-shell NAs adopting an icosahedron of 55 atoms (Ih$_{55}$), initially cleaved from Pt-bulk and then ionically relaxed. 
Ih, a Platonic solid, with twinning planes, made of 20 distorted tetrahedra sharing a common vertex, in such a way that only (111) facets are exposed. 
This structure is a commonly observed geometry for metallic nanoparticles and nanoalloys, especially at small sizes \cite{DiPaola2016,Sandoval2017,Baletto2002,Johnston2008}. 
Around the central atom, Ih shows an onion-shell motif, with a geometrical closure of the external shell after 12, 42, 92, 162, ... atoms. 
The initial 55-atoms core@shell configurations are obtained replacing the 13-innermost atoms by Ag, Ni and Pd atoms, respectively, keeping the outermost 
shell atoms as Pt ones. The systems are then ionically relaxed. Various inequivalent adsorption sites are considered for ethanol adsorption following 
the different coordination of the Pt-anchor site \cite{Davis2015,Batista2016,Silva2010,Piotrowski2012,Asara2016} and the relative position of the ethyl tail. 
Indeed, whether for Pd@Pt, the adsorption depends mainly on where the oxygen is anchored, the orientation of the ethyl groups is fundamental in Ag@Pt and Ni@Pt. 
Due to its central role, we introduce a new notation/nomeclature to distinguish the non-equivalent active sites reporting explicitly where the 
CH$_3$ and CH$_2$ are located relative to the NA. 

We name each site as [(oxygen position)]+[(CH$_3$ position)(CH$_2$ position)] and we use the labels E and V for edge and vertex Pt-atoms, respectively; 
and the tags {\it t, b, h} to identify whether the adsorption mode is atop, bridge, and hollow, respectively. For bridge and hollow, we list the relative 
position for all the Pt atoms involved. 

The oxygen atom of ethanol onto a metallic Ih$_{55}$ may lie on the top edge (tE), top vertex (tV), a bridge between edge and vertex (bEV), a bridge between two 
edge sites (bEE) and on hole site between a vertex and two edge sites (hVEE). Nonetheless, the latter as well as configurations where the C-C bond is 
radially oriented towards the NA, are energetically so unfavourable that turn to be unstable over all the considered NAs. These cases are not considered further.

For the ethyl group, a hydrogen contributes significantly to the adsorption energy when it is closer than 3.2 \AA{} from a Pt-atom. Interestingly, this is 
three-quarters of the bond length of an adsorbed hydrogen onto Pt(111) \cite{Papoian2000}. Table \ref{tab1} reports the distances (in \AA) between the 
Pt-atoms and the closest H-atom in CH$_3$ and CH$_2$ per each adsorption site considered. In light of the covalent bonding between the Pt-anchor and the 
ethanol O-atom, we consider the Pt-anchor exclusively bounded to the ethanol OH and not contributing to the weak bond of the CH-groups. Further, regarding 
the ethyl position on Pt-surface, three main adsorption sites are noticed. First, one surface Pt-atom can be closer to an ethanol H-atom over the evaluated 
sites, while other H-atoms are keeping farther. Then, two surface Pt-atoms can be at intermediate distances from an H-atom. Finally, some sites present 
much larger Pt-H distances for all H-atoms on CH$_2$ group.

Based on this, we define a bridge position whether the distances of H from the two underneath Pt-atoms are lower than 3.40 \AA{}, and
both lenghts by less than 0.4 \AA{}. The atop sites occur when one Pt-H distance in CH$_3$ or CH$_2$ is much closer compared to 
others (differing by more than 0.4 \AA{}), being around the range of the sum of H and Pt van der Waals radii, 2.95 \AA{}

For example, the label tE+bEEbEV refers to ethanol, OH+CH$_3$CH$_2$, positioned as O on top of a Pt-edge, CH$_2$ making a bridge 
with an edge and vertex Pt, and CH$_3$ sees a bridge between two edge Pt-atoms instead. This definition encloses all ethyl tail position 
on nanoparticles, except one (see table Table \ref{tab1}), where the lowest Pt-H distance is just 1\% greater than the H+Pt 
van der Waals radii. 

\begin{figure*}
  \includegraphics[width=0.75\textwidth,clip]{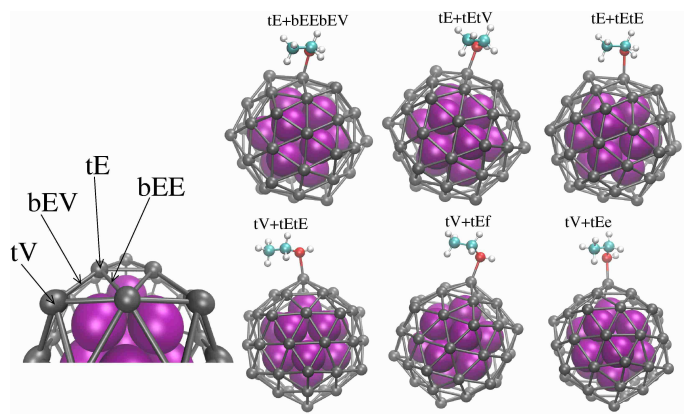}
\caption{Left snapshot introduces the nomenclature adopted to identify the $t$ (top), $b$ (bridge) and $E$ (edge), $V$ (vertex) 
sites for ethanol adsorption onto an Ih$_{55}$. Right panels show the six non-identical adsorption sites and their nomenclature 
reflecting the various orientations of the CH-groups with respect to the NA, as described in the text. Pt in silver, in purple 
the metallic core M, while red, cyan, white, stand for O, C and H, respectively.}
\label{fig1}       
\end{figure*}

Additionally, two cases present the CH$_2$ unbounded, with Pt-CH$_2$ distances greater than 3.8 \AA{}, while CH$_3$ reminds onto a tE site. 
This happens when the C-C bond is parallel to a (111) facet, tV+tEf; or when it lies along an edge, tV+tEe. In the end, we can 
distinguish six stable and non-identical adsorption sites for ethanol onto a Ih$_{55}$ NA, which are shown in Fig. \ref{fig1}. 

\begin{table}
\caption{Distances between surface Pt and the closest H-atom in CH$_3$ and CH$_2$ per each adsorption site and alloy. The label $d$ indicates the distance between an H atom and the Pt-anchor site.}
\label{tab1}       
\begin{tabular}{lll}
\hline\noalign{\smallskip}
  Site & CH$_{3}$ & CH$_{2}$ \\
\noalign{\smallskip}\hline\noalign{\smallskip}
      \multicolumn{3}{l}{Ag@Pt}\\
       tE+bEEbEV& 2.81/3.11$_d$/3.17 & 3.01/3.17/3.26$_d$ \\
       tE+tEtV  & 2.75/3.12$_d$/3.44 & 2.66/3.31$_d$/3.72 \\
       tE+tEtE  & 2.45/3.12$_d$/3.59 & 3.18/3.25$_d$/3.59 \\
       tV+tEtE  & 2.84/2.99$_d$/3.39 & 2.65/3.42$_d$/4.20 \\
       tV+tEf   & 2.97$_d$/3.06/3.88 & 3.34$_d$/4.73/4.09\\
       tV+tEe   & 2.69/3.02$_d$/4.02 & 3.31$_d$/3.83/4.32 \\
       \noalign{\smallskip}\hline\noalign{\smallskip}
       \multicolumn{3}{l}{Ni@Pt}\\
       tE+bEEbEV& 2.91/3.19$_d$/3.19 & 3.02/3.23$_d$/3.33 \\
       tE+tEtV  & 2.87/3.19$_d$/3.47 & 2.75/3.31$_d$/3.77 \\
       tE+tEtE  & 2.60/3.29$_d$/3.70 & 3.23$_d$/3.27/3.62 \\
       tV+tEtE  & 2.79/3.10$_d$/3.33 & 2.88/3.43$_d$/4.19 \\
       tV+tEf   & 3.06$_d$/3.20/3.95 & 3.47$_d$/4.23/4.79 \\
       tV+tEe   & 2.86/3.08$_d$/4.08 & 3.42$_d$/3.98/4.44 \\
       \noalign{\smallskip}\hline\noalign{\smallskip}
       \multicolumn{3}{l}{Pd@Pt}\\
       tE+bEEbEV& 2.92/3.09$_d$/3.16 & 2.99/3.27/3.24$_d$ \\
       tE+tEtV  & 2.74/3.14$_d$/3.48 & 2.65/3.28$_d$/3.68 \\
       tE+tEtE  & 2.50/3.30$_d$/3.45 & 3.08$_d$/3.24/3.69 \\
       tV+tEtE  & 2.72/3.09$_d$/3.43 & 2.99/3.10$_d$/4.09 \\
       tV+tEf   & 3.02/3.04$_d$/3.85 & 3.32$_d$/3.99/4.65 \\
       tV+tEe   & 2.66/3.04$_d$/3.97 & 3.27$_d$/3.76/4.24 \\
\noalign{\smallskip}\hline
\end{tabular}
\end{table}

The adsorption energy, E$_{ads}$, of ethanol on a given site is given by
\begin{equation}
    E_{ads}(site) = E_{NA+Et} - (E_{NA} + E_{Et}) \mbox{,}
\end{equation}
where $E_{NA+Et}$ is the total energy after the deposition of an ethanol molecule onto the selected site, while $E_{NA}$ and $E_{Et}$ are the total 
energies of the pristine NA and the molecule in the gas phase, respectively.


In addition to the energetic analysis, we present an extensive electronic and geometrical study to address the NA's core effects, 
and hence to suggest the most promising chemical composition. All together energetic, electronic, and geometrical analysis allow to 
characterise and unravel the role of the various adsorption sites; to eventually identify a new relationship linking geometrical properties
to the binding energy.

To access the electronic changes due to the ethanol adsorption, the charge redistribution is determined based on the charge difference for each atom $i$ before 
and after adsorption, 
\begin{equation}
\Delta q^{i} = q^{i}_{NA+Et} - q^{i}_{before} \mbox{,}
\end{equation}
where $q^{i}_{NA+Et}$ and $q^{i}_{before}$ are the Bader charges of the atom $i$, after/before adsorption, respectively. 

Additionally, we report the charge difference $\delta q^{i}$, between the atomic nominal valence charge, $q^{i}_{val}$, and its Bader charge after deposition. 
This provides a much-needed information on the charge transfer and the electrostatic effects. To address the metallic-core effects, we have determined the charge 
transfer from the ethanol, $\delta q^{Et}$, from the outer Pt-layer $\delta q^{shell}$, sub-surface $\delta q^{subshell}$, and the core atom, $\delta q^{core}$. 
This was calculated by summing the $\delta q^{i}$ contribution arising from atoms $i$ belonging to a certain subsystem,  for example $i \in {Et}$ or $i \in {Pt-shell}$,

To quantitatively characterise the adsorbate/NA interaction, we have monitor how the NA geometrical features vary after adsorption. Here, we propose the 
radial breathing, referring to the change of the radial position $r^{i \in NA}$ of the atoms $i$ within the NA with respect to the NA's centre of mass, 
before and after (NA+Et) ethanol adsorption.   
\begin{equation}
	\Delta r^{i \in NA}=r^{i}_{NA+Et}-r^{i}_{before} \mbox{,}
\end{equation}
Summing over each individual atomic contribution, we obtain the net radial distortion ND$_{NA} = \sum_{i \in NA} \Delta r^{i}$, 
taken with their sign while the absolute distortion, AD$_{NA}$ is $\sum_{i \in NA} |\Delta r^{i}|$. Again, we can distinguish 
between M-core and Pt-outermost shell, ND$_{subshell}$ and ND$_{shell}$, simply restricting the sum over a sub-system only. 
Similarly, it can be applied for AD$_{subshell}$ and AD$_{shell}$.

Finally to estimate the geometrical distortion within the ethanol molecule, we calculate the contraction/elongation of O-C and C-C 
bonds of the molecule in the gas phase and after chemisorption. Let $\Delta d_{O-C}$ and $\Delta d_{C-C}$ be the variations in those 
chemical bondings, where a positive sign will stand for an elongation, while a negative sign for a contraction with respect to the gas phase.

\section{Results and Discussion}
\label{rand}

\begin{figure*}
  \includegraphics[width=0.75\textwidth]{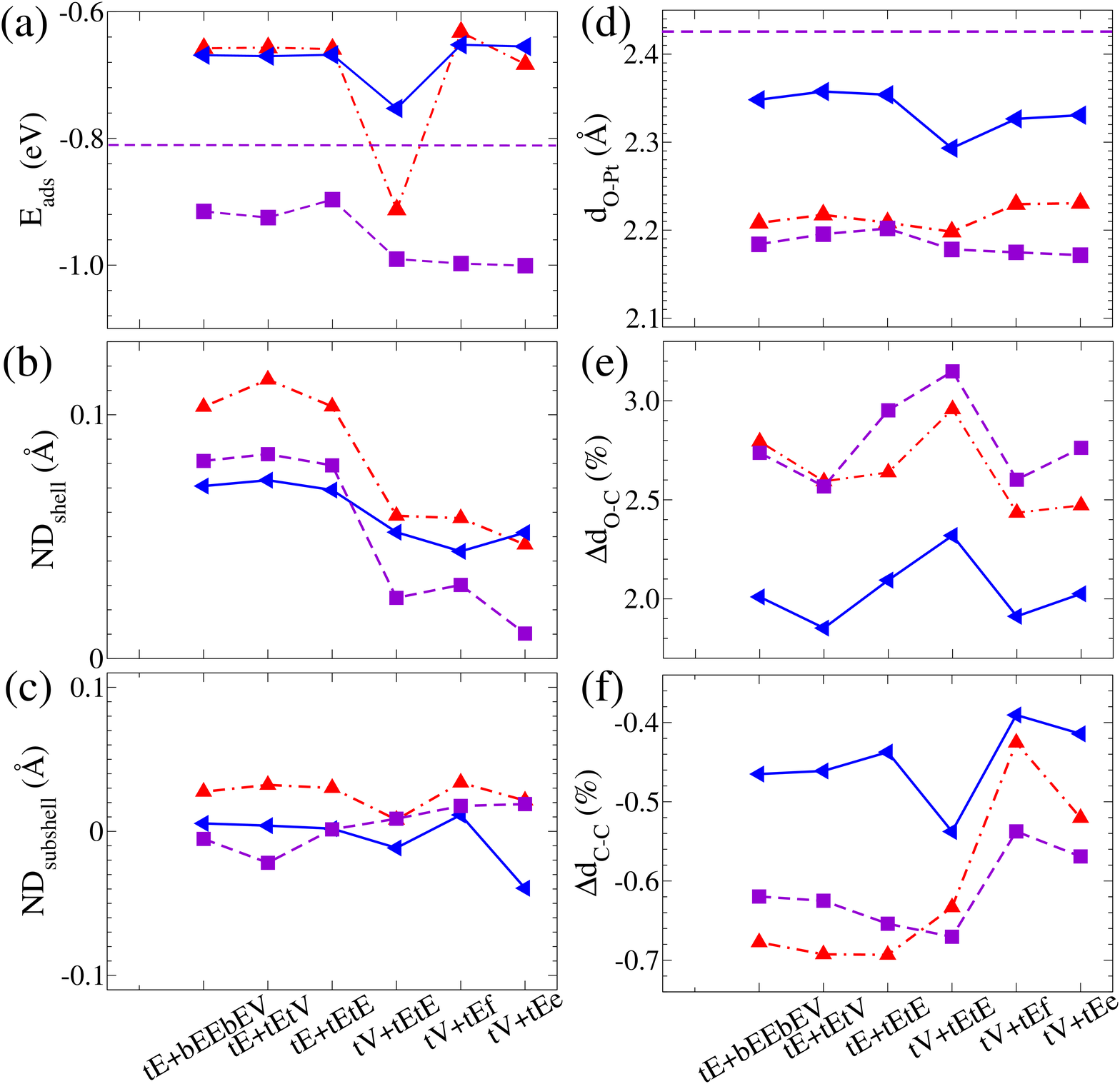}
\caption{(a) Adsorption energies, $E_{ads}$, on Ih$_{55}$ Ni@Pt (blue right-triangle, full-line), Ag@Pt (red up-triangle, dashed-line), 
and Pd@Pt (purple square, dashed line). Purple horizontal line refers to Pd/Pt nanofilm taken from Ref. \cite{Pereira2014} whereas, the E$_{ads}$ on Ni/Pt nanofilm (-0.2 eV),
from \cite{Skoplyak2008}, lies well up the scale of our graph; (b) Net distortion on shell, ND$_{Pt}$ and (c) core ND$_{core}$; 
(d) bond lenght O-Pt-docking site, d$_{O-Pt}$; (e) percentage contraction of $\Delta d_{O-C}$ and (f) $\Delta d_{C-C}$ after ethanol adsorption. 
Lines are only to guide the eye vision.}
\label{fig2}       
\end{figure*}

Fig. \ref{fig2} summarises the adsorption properties, both energetic and geometrical, of an ethanol molecule, the three Pt-shell M-core 
systems and the six adsorption sites listed in Fig. \ref{fig1}. Let us first comment on the binding energy, reported in Fig.~\ref{fig2}(a).
For Pd@Pt and Ni@Pt systems, there is an enhancement up to 0.19 eV and 0.52 eV, respectively,relative to the equivalent 
nanofilms \cite{Pereira2014,Skoplyak2008}. Nonetheless, it is worth to note that the adsorption on Ni@Pt surfaces did not include any 
van der Waals correction \cite{Skoplyak2008}, which might play a role. At the best of our knowledge, there are no data for Ag@Pt to compare with.

On Pd@Pt, the average E$_{ads}$ is -0.95 eV with a difference as small as 0.08 eV between tE and tV sites, and a weak if not negligible 
dependence on the orientation of the CH-tail. We note that this value is also higher than the binding energy calculated on a Ih$_{13}$ \cite{ZB2016}.
Although out of the aim of this work, this seems to indicate a peculiar size dependence of ethanol adsorption energy. On Ag@Pt, the average 
distribution is of -0.70 eV and it shows a strong dependence on the CH-groups orientation, with five sites almost lying at -0.67 eV, and the 
tV+tEtE as low as -0.91 eV, similar to the values of Pd@Pt when oxygen is top-edge. Ni@Pt presents a similar behaviour to Ag@Pt, with an 
almost flat binding energy around -0.65 eV and a drop at the tV+tEtE sites of about 0.1 eV.

The Pd@Pt presents the shorter d$_{O-Pt}$ values, contracted by ~10$\%$ with respect to their surface equivalent. This could suggest that Ih$_{55}$ 
is a good candidate for EOR, as a contraction of d$_{O-Pt}$ is usually associated with an improvement of the catalytic reaction \cite{Alcala2003}.
Notably, the d$_{O-Pt}$ values on Fig. \ref{fig2}(d) seems correlated with the E$_{ads}$ on Fig. \ref{fig2}(a). However, the shorter d$_{O-Pt}$ 
values occur on tV+tEtE site on Ag@Pt and Ni@Pt, and on tV+tEe site on Pd@Pt. These adsorption sites are the most stable for each nanoparticle, 
indicating that the d$_{O-Pt}$ is relevant to the system energetics, although not exclusively. 

A step forward to the understanding of the peculiar behaviour of Ag@Pt and Ni@Pt is achieved taking into account the geometrical
distortion induced by ethanol onto the shell and subshell of the clusters, Fig. \ref{fig2}(b) and Fig. \ref{fig2}(c), respectively. 
The radial breathing of the structures, not necessarily symmetric, especially on the outer shell atoms, can represent a relevant 
topic for EOR. Notably, each nanoalloy presents a particular breathing characteristic upon ethanol adsorption. The ND$_{shell}$ are 
site dependent and usually, the main cluster distortion happens for tE adsorption and in non-mismatched nanoalloys. Although preserving 
the Ih symmetry, the ND$_{shell}$ increase is higher when ethanol is adsorbed on edges rather the on vertices, and this effect is
more pronounced on Ag@Pt and Pd@Pt, than Pd@Pt, probably due to their different mismatch. These results indicate a dependence between 
the coordination of surface atoms and the ND$_{shell}$. A few rearrangements occur in the subshell, as verified in Fig. \ref{fig2}(c).

\begin{figure*}
  \includegraphics[width=0.75\textwidth]{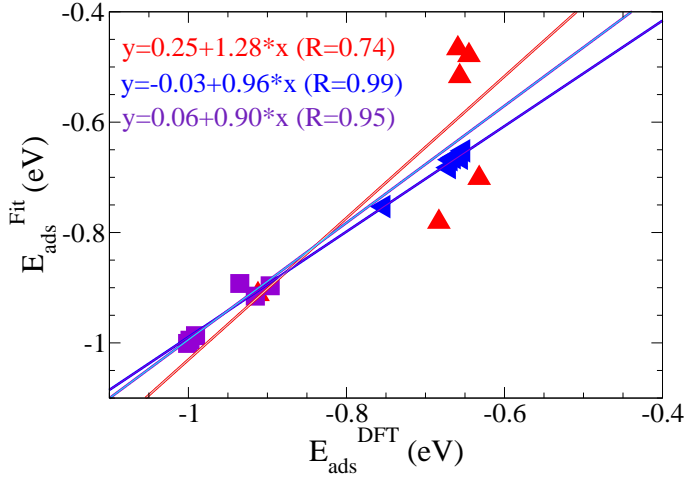}
\caption{Ethanol adsorption energy obtained with DFT and the adjusted equation (E$_{ads}^{Fit}$) for each adsorption site,
with Ag@Pt as red triangle-up, Ni@Pt as blue triangle-left and Pd@Pt as purple squares.
The lines correspond to linear fittings of the obtained data, for each alloy.
Relevant data of fittings are presented in Table \ref{tab3} and discussed along with the text.}
\label{fig4}       
\end{figure*}

A dependence on both d$_{O-Pt}$ and ND$_{shell}$ in hence needed to estimate the binding energy on Pt-shell M-core nanoalloys
also with a significant lattice mismatch. This can be expressed by a linear relationship
\begin{equation}
E_{ads}^{Fit} = \alpha d_{O-Pt} + \beta ND_{shell} + \gamma \mbox{,}
\label{eq3}
\end{equation} 
where $\alpha$, $\beta$ and $\gamma$ represent the adjusted constants, with values presented in Table \ref{tab3}.
To evaluate the quality of Eq. \ref{eq3} and Table \ref{tab3} to reproduce the DFT data, Fig. \ref{fig4} presents the
fitted values, E$_{ads}^{Fit}$, as a function of the DFT data, for each site and nanoalloy, including the linear fitting of each curve.
The Ni@Pt and Pd@Pt fittings represent the E$_{ads}$ values with notable accuracy, confirmed by the R close to one, 
as can seems in Fig. \ref{fig4}. Although the E$_{ads}^{Fit}$ of Ag@Pt follows the E$_{ads}$ trend for the most stable site, 
the fitting presents a poor quality, indicated by the obtained low R (0.74).
Eq. \ref{eq3} reproduces with the energetics of the most stable site of each core composition, indicating that d$_{O-Pt}$ and 
ND$_{shell}$, uniquely, are the key features that control the most stable position of ethanol on nanoparticles.
The E$_{ads}^{Fit}$ and DFT E$_{ads}$ values at the most stable site results in an error of approximately 1\% on Ag@Pt, and less 
than 1\% on Ni@Pt and Pd@Pt.
Further, the signal of the $\beta$ coefficient on Eq. \ref{eq3} varies according the nanoalloy (Table \ref{tab3}),
evidencing the effects of the chemical composition and strain effects on the system energetics.
Notably, $\beta$ is positive on Pd and Ag, and negative on Ni-core. This seems to be related to the induced strain depending 
on the mismatch. In particular, the van der Waals radii of Ag, Ni, Pd and Pt are 2.13, 1.94, 2.05 and 2.06 \AA{} \cite{Batsanov2009}. 
Comparing the van der Waals radii of core elements with one Pt one, a greater mismatch is noted between Ni and Pt, indicating 
that Eq. \ref{eq3} reproduces part of the strain effects on E$_{ads}$, although not completely.

The data obtained with Eq. \ref{eq3} are less able to reproduce the E$_{ads}$ far apart from the minimum energy configuration, 
specifically on Ag@Pt, and also Pd@Pt to a lesser extent. This appears to be related to the dispersion interaction between 
the ethanol ethyl group and surface atoms, enhanced due to the strained surface of Ag-core.
From Table \ref{tab1}, the CH$_2$ on a tV site reduces the distance from the Pt-surface, whereas, the CH$_3$ on tE is closer to
Pt-surface on Ag@Pt, specilly on tE+tEtV and tE+tEtE sites, and also noted on tE+tEf and tE+tEe ones.

Moving to the effects of the adsorbed molecule, from panels (e) and (f) of Fig. \ref{fig2}, we systematically observe an elongation 
of the O-C, with a peak on the tV+tEtE site, and a contraction of the C-C bond. Nonetheless, $\Delta d_{O-C}$ is larger in Pd@Pt, and less in Ni@Pt. 
We would like to comment on the different role played by the CH$_{x}$ ($x$=2,3) groups onto the cluster: 
CH$_3$ on tE shows a shorter elongation of the O-C bond. On the other hand, when both CH$_2$ and CH$_3$ are on top of a Pt-edge, $\Delta d_{O-C}$ peaks. 
Fig. \ref{fig2}(f) shows that the C-C contraction is dependent on the core composition, namely when oxygen is adsorbed on 
top of vertex sites, the shorter C-C bond occurs on Pd@Pt core, whereas, on edge sites, the shorter C-C bonds occurs on the 
Ag@Pt nanoparticle. This information is important for EOR, since the elongation on the atomic distances points towards a 
favourable bond scission \cite{Alcala2003}.

\begin{table}
\caption{Values of the adjusted constants regarding Eq. \ref{eq3}, for each evaluated nanoalloy.}
\label{tab3}       
\begin{tabular}{llll}
\hline\noalign{\smallskip}
Nanoalloy & $\alpha$ (eV/\AA{}) & $\beta$ (eV/\AA{}) & $\gamma$ (eV) \\
\noalign{\smallskip}\hline\noalign{\smallskip}
       Ag@Pt & 3.76  & 6.82  & -9.57  \\
       Ni@Pt & 2.32  & -3.01 & -5.92  \\
       Pd@Pt & 2.45  & 0.78  & -6.37  \\ 
\noalign{\smallskip}\hline
\end{tabular}
\end{table}

Let us now discuss the electronic contributions in the case of ethanol onto a tE+tEtV, tV+tEtE and tV+tEe sites, which is representative 
of a tE site, and tV ones are the best for both Ag@Pt and Ni@Pt, and Pd@Pt, respectively. Table \ref{tab2} presents the charge difference of 
the Pt-anchor ($\Delta q^{Pt}$) site and the ethanol oxygen $\Delta q^{O}$ before and after ethanol adsorption, and the charge transfer 
after adsorption ($\delta q^{i}$) of the Pt-anchor site ($\delta q^{Pt}$), ethanol oxygen ($\delta q^{O}$), and each subsystem, namely 
ethanol molecule (Et) and metallic inner core (subshell and core) and Pt-external layer (shell). These data provide relevant information about the systems. 
Interestingly, charge difference, $\Delta q^{i}$, shows that both the ethanol-oxygen and the Pt-anchor lose electrons after the adsorption 
on all nanoalloys. Although this charge redistribution depends on the chemical core.

The charge transfer, $\delta q^{i}$, shows that the sub-surface layer is always positively charged, while the extra electrons migrate mainly 
towards the external shell. This effect is more significant in Ni@Pt, and less in Pd@Pt. As a result, the Pt-anchor is less positively charged 
on Ni@Pt (only 0.02 electrons), explaining the longer d$_{O-Pt}$ and weaker E$_{ads}$ obtained on Ni@Pt, compared to the other compositions seen here. 
This is consistent with experimental data, where the EOR on a Pt-Ni/$\delta$Al$_2$O$_3$ surface occurs only at high-temperature \cite{Orucu2008}.
At the same time, we note that the Pd-core associates significant changes of intra-ethanol bonds and a significant charge transfer from the molecule 
towards the nanoparticle and overall ethanol as stronger bounded to the cluster.

\begin{table}
\caption{Charge transfer upon adsoprtion (relative to valence charge) $\delta q^{i}$, and charge difference (before and after adsorption) $\Delta q^{i}$, for 
    Pt-anchor, ethanol oxygen, ethanol molecule (Et), nanoparticle shell, subshell, and core-atom. Positive values indicate a gain of charge, whereas, subsystems with 
    negative values donates charge, in the unit of electrons.}
\label{tab2}       
\begin{tabular}{lllllllll}
\hline\noalign{\smallskip}
System & $\delta q^{Pt}$ & $\delta q^{O}$ & $\Delta q^{Pt}$ & $\Delta q^{O}$ & $\delta q^{Et}$ & $\delta q^{shell}$ & $\delta q^{subshell}$ & $\delta q^{core}$ \\
\noalign{\smallskip}\hline\noalign{\smallskip}
      \multicolumn{9}{l}{Ag@Pt} \\
       tE+tEtV & -0.12 & 1.54 & -0.17 & -0.21 & -0.14 & 3.39 & -3.23 & -0.02 \\
       tV+tEtE & -0.07 & 1.56 & -0.18 & -0.20 & -0.12 & 3.41 & -3.26 & -0.02 \\
       tV+tEe  & -0.05 & 1.56 & -0.17 & -0.19 & -0.14 & 3.41 & -3.25 & -0.02 \\
       \noalign{\smallskip}\hline\noalign{\smallskip}
       \multicolumn{9}{l}{Ni@Pt}\\
       tE+tEtV & -0.05 & 1.59 & -0.12 & -0.16 & -0.10 & 4.22 & -4.19 & 0.07 \\
       tV+tEtE & -0.02 & 1.58 & -0.16 & -0.17 & -0.11 & 4.24 & -4.20 & 0.07 \\
       tV+tEe  &  0.02 & 1.60 & -0.12 & -0.15 & -0.11 & 4.24 & -4.20 & 0.07 \\
       \noalign{\smallskip}\hline\noalign{\smallskip}
       \multicolumn{9}{l}{Pd@Pt} \\
       tE+tEtV & -0.15 & 1.55 & -0.18 & -0.21 & -0.13 & 2.64 & -2.54 & 0.03 \\
       tV+tEtE & -0.07 & 1.54 & -0.18 & -0.22 & -0.14 & 2.63 & -2.55 & 0.06 \\
       tV+tEe  & -0.07 & 1.54 & -0.19 & -0.21 & -0.14 & 2.64 & -2.43 & 0.06 \\
\noalign{\smallskip}\hline
\end{tabular}
\end{table}

\section{Conclusion}
Concluding, on the view of both the geometrical and electronic analysis, it seems that the Ag@Pt and Pd@Pt Ih$_{55}$ 
present some promising features towards ethanol oxidation reaction. Ethanol shows a similar charging transfer upon 
adsorption on both nanoalloys. The Pd@Pt presents the shortest d$_{O-Pt}$, and the largest O-C bond when ethanol
is in atop position over a five-fold vertex, whereas Ag@Pt enlarges the O-C bond when the molecule is atop but on Pt-edge. 
Those data are explained in terms of the charging analysis, where the charge transfer of the Pt-anchor is similar to all 
the considered nanosystems, with 0.12 and 0.14 electrons on Ag@Pt and Pd@Pt, respectively. A model was proposed to 
describe the ethanol stability on each site, as a function of d$_{O-Pt}$ and the net radial distortion in the external 
Pt-shell, which is able to describe with accuracy the obtained data for Ni@Pt and Pd@Pt, and Ag@Pt in a lesser account, 
but explaining the most stable configuration. Finally, the Ag@Pt and Pd@Pt Ih$_{55}$ seems to be the most cost-effective 
material to EOR, compared to pure Pt, whereas the Ni@Pt shows much less attractive adsorption properties for this chemical reaction.

\begin{acknowledgements}
FB thanks the financial support by the U.K Engineering and Physical Sciences Research Council (EPSRC), under Grants No. EP$/GO03146/1$ and No. \linebreak \mbox{EP$/J010812/1$.}
VAR and CRM thanks the support of CNPq - Brazilian Federal Agency for Scientific and Technological Development - Brazil - along this work.
All the authors thanks the KCL/FAPESP agreement, and the computational support of CENAPAD-SP, UTFPR-CP, and Sdumont, in Brazil.
The authors thank the financial support offered by Royal Society under the project number RG120207 and FAPESP Grant No. 17/02317-2.
\end{acknowledgements}

\end{document}